\documentclass{revtex4-2}
\usepackage{lineno}
\usepackage[utf8]{inputenc}
\usepackage{amsbsy}
\usepackage{amsopn}
\usepackage{amstext}
\usepackage{amsmath,amsthm,amsfonts,amssymb}
\usepackage[mathcal]{eucal}
\usepackage{mathrsfs}
\usepackage{braket}
\usepackage[english]{babel}
\usepackage{color}
\usepackage{esint}
\usepackage{booktabs}
\usepackage{hologo}
\usepackage{graphicx}
\usepackage{float}
\usepackage{units}
\usepackage{textcomp}
\usepackage{orcidlink}

\DeclareGraphicsExtensions{.png,PNG,.pdf,.PDF}
\usepackage{hyperref}
\usepackage{slashed}
\newcommand{\ie}{\begin{equation}}
\newcommand{\fe}{\end{equation}}
\newcommand{\se}{\begin{eqnarray}}
\newcommand{\ff}{\end{eqnarray}}
\begin{document}
\title{Dirac fermions under rainbow gravity effects in the Bonnor-Melvin-Lambda spacetime}
\author{R. R. S. Oliveira\,\orcidlink{0000-0002-6346-0720}}
\email{rubensrso@fisica.ufc.br}
\affiliation{Departamento de F\'isica, Universidade Federal do Cear\'a (UFC), Campus do Pici, C.P. 6030, Fortaleza, CE, 60455-760, Brazil}


\date{\today}

\begin{abstract}
In this paper, we study the relativistic energy spectrum for Dirac fermions under rainbow gravity effects in the $(3+1)$-dimensional Bonnor-Melvin-Lambda spacetime, where we work with the curved Dirac equation in cylindrical coordinates. Using the tetrads formalism of General Relativity and considering a first-order approximation for the trigonometric functions, we obtain a Bessel equation. To solve this differential equation, we also consider a region where a hard-wall confining potential is present (i.e., some finite distance where the radial wave function is null). In other words, we define a second boundary condition (Dirichlet boundary condition) to achieve the quantization of the energy. Consequently, we obtain the spectrum for a fermion/antifermion, which is quantized in terms of quantum numbers $n$, $m_j$ and $m_s$, where $n$ is the radial quantum number, $m_j$ is the total magnetic quantum number, $m_s$ is the spin magnetic quantum number, and explicitly depends on the rainbow functions $F(\xi)$ and $G(\xi)$, curvature parameter $\alpha$, cosmological constant $\Lambda$, fixed radius $r_0$, and on the rest energy $m_0$, and $z$-momentum $p_z$. So, analyzing this spectrum according to the values of $m_j$ and $m_s$, we see that for $m_j>0$ with $m_s=-1/2$ (positive angular momentum and spin down), and for $m_j<0$ with $m_s=+1/2$ (negative angular momentum and spin up), the spectrum is the same. Besides, we graphically analyze in detail the behavior of the spectrum for the three scenarios of rainbow gravity as a function of $\Lambda$, $r_0$, and $\alpha$ for three different values of $n$ (ground state and the first two excited states).
\end{abstract}

\maketitle

\section{Introduction}

The so-called Dirac fermions are spin-1/2 massive particles modeled by the Dirac equation (DE), which is a relativistic wave equation derived by P. A. M. Dirac in 1928, and is one of the fundamental equations of the Relativistic Quantum Mechanics (RQM) \cite{Dirac1,Dirac2,Greiner,Strange}. Therefore, spin-1/2 massive particles (or simply spin-1/2 particles) are also often so-called Dirac particles, such as the electrons, protons, neutrons, quarks, muons, taus, and possibly the neutrinos \cite{Studenikin,Lesgourgues}. In addition to the DE naturally explaining (or proving) the spin, helicity, and chirality of the spin-1/2 particles, predict that for each of these particles, there are their respective antiparticles (e.g., positrons, antiprotons, antineutrons,...) \cite{Dirac1,Dirac2,Greiner,Strange,Studenikin,Lesgourgues,Martin,griffiths}. In particular, the DE has many applications, for example, can be applied to model or study the Dirac oscillator (a relativistic quantum harmonic oscillator for Dirac fermions) \cite{Moshinsky,Franco,Oliveira0}, Aharonov-Bohm and Aharonov-Casher effects \cite{Aharonov,Hagen1,Hagen2,Oliveira1}, Aharonov-Bohm-Coulomb system \cite{Oliveira2,Oliveira3}, quantum rings \cite{Oliveira4}, ultracold and cold atoms \cite{Zhang2012}, trapped ions \cite{Lamata}, quantum Hall effect \cite{Schakel}, quantum computing \cite{Huerta}, quantum phase transitions \cite{Bermudez1}, mesoscopic superposition states \cite{Bermudez2}, quark models \cite{Becirevic}, Dirac materials (e.g., graphene, fullerenes, metamaterials, semimetals, and carbon nanotubes) \cite{Novoselov,Gonzalez,McCann,Ahrens,Armitage}, etc. Recently, the DE has been used to investigate the Dirac oscillator \cite{SousaOliveira,Oliveira5}, quantum Hall effect \cite{Oliveira6,Oliveira,Oliveira7}, quantum interference and entanglement \cite{Ning}, electron-nucleus scattering \cite{Jakubassa}, non-Hermitian physics and strong correlation \cite{Yu}, spinning conical Gödel-type spacetime \cite{ARXIV2}, etc.

In General Relativity (GR), the so-called rainbow gravity is a semi-classical theory to probe high-energy phenomena of quantum gravity via the relativistic dispersion relation with higher-order terms originated by the so-called rainbow functions (thus implying a violation of Lorentz symmetry) \cite{Magueijo1,Magueijo2,Magueijo3,Amelino1,Amelino2,Bezerra2019,Bakke2018}. In particular, these rainbow functions depend on the ratio between the energy of a test particle (bosons or fermions) and on the Planck energy (i.e., $\xi=E/E_P\leq 1$). However, how in rainbow gravity there are two quantities that are observer-independent (speed of light and the Planck energy), meaning that this gravity has its origin in so-called Doubly General Relativity (DGR), that is the Doubly Special Relativity (DSR) applied (generalized) to GR (or curved spacetimes) \cite{Magueijo1,Magueijo2,Magueijo3,Amelino1,Amelino2,Bezerra2019,Bakke2018}. In that way, to build the framework of rainbow gravity, the line element (or metric) of spacetime has also to be energy-dependent (i.e., dependent of $\xi$) \cite{Magueijo1,Magueijo2,Magueijo3,Amelino1,Amelino2,Bezerra2019,Bakke2018}. Besides, unlike the usual GR, where gravity acts equally regardless of the particle's energy, rainbow gravity acts differently on particles with different energies, i.e., for each particle's energy value, it feels different gravity levels. Indeed, this effect is very small for objects like the Earth, however, it should become significant for objects like black holes. For example, one ``simple way" to detect such effects would be via ultra-high energy cosmic rays or TeV photons and neutrinos from gamma-ray-bursts (GRB), where the relativistic dispersion relation should acquire corrections (modifications) due to the parameter $\xi$ \cite{Magueijo2,AmelinoCamelia1,AmelinoCamelia2,Jacob,Zhang,Nilsson}. Thus, after being introduced in the literature, rainbow gravity gained a lot of attention and also many interesting applications, such as in neutron stars \cite{Hendi1}, black holes \cite{Hendi2,Hendi3,Ali1}, wormholes \cite{Amirabi}, Friedmann-Robertson-Walker (FRW) cosmology \cite{Hendi,Awad}, Casimir effect \cite{Bezerra2017}, Klein–Gordon oscillator \cite{Montigny1}, bosonic fields \cite{Faizuddin}, Landau levels \cite{ARXIV1}, etc.

Furthermore, a solution or model of particular relevance in GR is the so-called Bonnor-Melvin spacetime (or Bonnor-Melvin universe), which is an exact solution of Einstein-Maxwell equations that describes a static and cylindrically symmetric (electro)magnetic field immersed in its own gravitational field (where the magnetic field is aligned with the symmetry axis) \cite{Bonnor,Melvin,Zofka,Cataldo,Mazharimousavi,Ger}. That is, the Bonnor-Melvin spacetime (can also be called Bonnor-Melvin magnetic spacetime) describes the gravitational field generated by an axial magnetic field permeating the whole spacetime due to the azimuthal current on the surface of a coaxial cylinder enveloping part of the spacetime \cite{Vesel}. Recently, has been proposed in the literature a generalization of the Bonnor-Melvin spacetime, the so-called Bonnor-Melvin-Lambda spacetime, which is the Bonnor-Melvin spacetime with a cosmological constant \cite{Zofka}. In this generalization, where a non-vanishing cosmological constant is considered, the spacetime is again cylindrically symmetric and static; however, unlike the original spacetime, it truly represents a homogeneous magnetic field \cite{Zofka}. In addition, this generalization also admits a deficit angle due to the presence of a (curvature) parameter $\alpha$ in the line element (or metric) and, therefore, forming the spacetime’s only singularity (i.e., a conical or cone-like singularity) \cite{Zofka}. In particular, the Bonnor-Melvin-Lambda spacetime has been used for studying the Klein-Gordon oscillator \cite{Ahmed1}, and the Duffin-Kemmer-Petiau oscillator \cite{Ahmed2}, while the Bonnor-Melvin-Lambda spacetime with rainbow gravity effects has been used for studying scalar particles through the Klein-Gordon equation \cite{Ahmed3}.

This paper has as its goal to study the relativistic energy spectrum (high energy spectrum) for Dirac fermions under rainbow gravity effects in the $(3+1)$-dimensional Bonnor-Melvin-Lambda spacetime. For this, we work with the curved DE (or modified DE by rainbow gravity) in cylindrical coordinates, where we use the tetrads formalism of GR to obtain the (bound-state) solutions of this equation. Indeed, we use this formalism because it is a great tool for studying fermions in curved spacetimes. Furthermore, to simplify the calculations, we consider a first-order approximation for the trigonometric functions in the second-order differential equation generated from DE. Subsequently, we also consider a region where a hard-wall confining potential is present, i.e., some finite distance where the radial wave function (spatial part of the Dirac spinor) is null/zero (i.e., must disappear). In other words, we define a second boundary condition (Dirichlet boundary condition) to achieve the quantization of the energy. Now, with respect to rainbow functions, we choose three different pairs of rainbow functions for $F(\xi)$ and $G(\xi)$ based on phenomenological motivations (and also the most worked in the literature). In this way, we work in three different scenarios (or three different cases) of rainbow gravity. Explicitly, these three pairs of rainbow functions are given by:
\begin{itemize}
\item First Scenario \cite{ARXIV1,Amelino1,Amelino2,Bezerra2019,Hendi2,Hendi3,Hendi,Hendi1,Faizuddin}:
\begin{equation}\label{f1}
F(\xi)=1, \ \ G(\xi)=\sqrt{1-\eta\xi},
\end{equation}
which is compatible with results from Loop Quantum Gravity and noncommutative spacetime, and $\eta>0$ is a dimensionless constant parameter, and $\xi$ is sometimes so-called the rainbow parameter.
\item Second Scenario \cite{ARXIV1,AmelinoCamelia1,Bezerra2019,Hendi2,Hendi3,Amirabi,Bezerra2017,Hendi,Awad,Hendi1,Faizuddin}:
\begin{equation}\label{f2}
F(\xi)=\frac{e^{\eta\xi}-1}{\eta\xi}, \ \ G(\xi)=1,
\end{equation}
which is compatible with the hard spectra from gamma-ray bursters at cosmological distances.
\item Third Scenario \cite{ARXIV1,Magueijo1,Bezerra2019,Magueijo3,Hendi2,Hendi3,Amirabi,Bezerra2017,Hendi,Awad,Hendi1,Faizuddin}:
\begin{equation}\label{f3}
F(\xi)=G(\xi)=\frac{1}{1-\eta\xi},
\end{equation}
which produces a constant speed of light, and might
solve the horizon problem.
\end{itemize}

The structure of this paper is organized as follows. In Sect. \ref{sec2}, we introduce the curved DE under rainbow gravity effects in the Bonnor-Melvin-Lambda spacetime. Using the tetrads formalism, we obtain a second-order differential equation for one of the components of the Dirac spinor. In Sect. \ref{sec3}, we consider a first-order approximation for the trigonometric functions and we obtain a Bessel equation, where we solve this differential equation through a region where a hard-wall confining potential is present, i.e., some finite distance (fixed radius) where the radial wave function is null. In other words, we define a second boundary condition (Dirichlet boundary condition) to achieve the quantization of the energy. Consequently, we obtain the relativistic energy spectrum for a fermion/antifermion (or particle/antiparticle), where we discuss some characteristics of this spectrum. Subsequently, we graphically analyze the behavior of the spectrum for the three scenarios of rainbow gravity as a function of the cosmological constant $\Lambda$, fixed radius $r_0$, and of the curvature parameter $\alpha$ for three different values of the radial quantum number $n$ (i.e., ground state, first excited state, and second excited state). In Sect. \ref{sec4}, we present our conclusions. Here, we use the natural units $(\hslash=c=G=1)$ and the spacetime with signature $(+,-,-,-)$.

\section{The Dirac equation under rainbow gravity effects in the Bonnor-Melvin-Lambda spacetime \label{sec2}}

The $(3+1)$-dimensional DE in an arbitrary curved spacetime is given by the following expression (in cylindrical coordinates) \cite{Oliveira7,Oliveira,ARXIV1,Lawrie}
\begin{equation}\label{dirac1}
(i\gamma^\mu(x)\nabla_\mu(x)-m_0)\psi(t,r,\theta,z)=0, \ \ (\mu=t,r,\theta,z),
\end{equation}
where $\gamma^{\mu}(x)=e^\mu_{\ a}(x)\gamma^a$ are the curved gamma matrices, which satisfy
the anticommutation relation of the covariant Clifford algebra: $\{\gamma^{\mu}(x),\gamma^{\nu}(x)\}=2g^{\mu\nu}(x)$, being $g^{\mu\nu}(x)$ the curved metric tensor (curved metric), $\gamma^a=(\gamma^0,\gamma^i)=\eta^{ab}\gamma_b$ are the usual or flat gamma matrices (in Cartesian coordinates), $\eta^{ab}=\eta_{ab}=$diag$(+1,-1,-1,-1)$ is the flat Minkokski metric, $e^\mu_{\ a}(x)$ are the tetrads, $\nabla_\mu(x)=\partial_\mu+\Gamma(x)$ is the covariant derivative, being $\partial_\mu=(\partial_t,\partial_r,\partial_\theta,\partial_z)$ the usual partial derivatives, $\Gamma_\mu(x)=-\frac{i}{4}\omega_{ab\mu}(x)\sigma^{ab}$ is the spinorial connection (spinor affine connection), being $\omega_{ab\mu}(x)$ the spin connection, $\sigma^{ab}=\frac{i}{2}[\gamma^a,\gamma^b]$ is a flat antisymmetric tensor, $\psi=e^{\frac{i\theta\Sigma_3}{2}}\Psi_D$ is the four-component curvilinear spinor, where $\Psi_D\in\mathbb{C}^4$ is the original Dirac spinor (four-element column vector), and $m_0>0$ is the rest mass of the Dirac fermion, respectively. Here, we use the Latin indices $(a, b, c, \ldots)$ to label the local coordinates system (local reference frame or the Minkowski spacetime) and the Greek indices $(\mu, \nu, \alpha, \ldots)$ to label the general coordinates system (general reference frame or the curved spacetime).

Explicitly, we can rewrite Eq. \eqref{dirac1} as follows
\begin{equation}\label{dirac2}
\{i\gamma^t(x)\partial_t+i\gamma^r(x)\partial_r+i\gamma^\theta(x)\partial_\theta+i\gamma^z(x)\partial_z-m_0+i[\gamma^t(x)\Gamma_t(x)+\gamma^{r}(x)\Gamma_{r}(x)+\gamma^{\theta}(x)\Gamma_{\theta}(x)+\gamma^{z}(x)\Gamma_{z}(x)]\}\psi=0,
\end{equation}
where the flat gamma matrices ($4\times 4$ matrices) are given as follows (in Dirac standard representation)
\begin{equation}\label{Diracmatrices}
 \gamma^0=\left(
    \begin{array}{cc}
      1 & \ 0 \\
      0 & -1 \\
    \end{array}
  \right), \ \ \gamma^i=\left(
    \begin{array}{cc}
      0\ &  \sigma^i \\
      -\sigma^i\ & 0 \\
    \end{array}
  \right), \ \ (i=1,2,3=x,y,z),
\end{equation}
where the Pauli matrices ($2\times 2$ matrices) given by
\begin{equation}\label{Paulimatrices}
\sigma_1=\left(
    \begin{array}{cc}
      0\ &  1 \\
      1\ & 0 \\
    \end{array}
  \right), \ \  \sigma_2=\left(
    \begin{array}{cc}
      0 & -i  \\
      i & \ 0 \\
    \end{array}
  \right), \ \  \sigma_3=\left(
    \begin{array}{cc}
      1 & \ 0 \\
      0 & -1 \\
    \end{array}
  \right).
\end{equation}

Besides, the spin connection (an antisymmetric tensor) is defined in the following form (torsion-free) \cite{Lawrie}
\begin{equation}\label{spinconnection}
\omega_{ab\mu}(x)=-\omega_{ba\mu}(x)=\eta_{ac}e^c_{\ \nu}(x)\left[e^\sigma_{\ b}(x)\Gamma^\nu_{\ \mu\sigma}(x)+\partial_\mu e^\nu_{\ b}(x)\right], 
\end{equation}
where $\Gamma^\nu_{\ \mu\sigma}(x)$ are the Christoffel symbols of the second type (a symmetric tensor), given by
\begin{equation}\label{Christoffel}
\Gamma^\nu_{\ \mu\sigma}(x)=\frac{1}{2}g^{\nu\lambda}(x)\left[\partial_\mu g_{\lambda\sigma}(x)+\partial_\sigma g_{\lambda\mu}(x)-\partial_\lambda g_{\mu\sigma}(x)\right], 
\end{equation}
where the tetrads and their inverses must satisfy the following relations
\begin{eqnarray}\label{tetrads}
&& g_{\mu\nu}(x)=e^a_{\ \mu}(x)e^b_{\ \nu}(x)\eta_{ab},
\nonumber\\
&& g^{\mu\nu}(x)=e^\mu_{\ a}(x)e^\nu_{\ b}(x)\eta^{ab},
\nonumber\\
&& g^{\mu\sigma}(x)g_{\nu\sigma}(x)=\delta^\mu_{\ \nu}=e^a_{\ \nu}(x)e^\mu_{\ a}(x),
\end{eqnarray}
as well as
\begin{eqnarray}\label{metric3}
&& \eta_{ab}=e^\mu_{\ a}(x)e^\nu_{\ b}(x)g_{\mu\nu}(x),
\nonumber\\
&& \eta^{ab}=e^a_{\ \mu}(x)e^b_{\ \nu}(x)g^{\mu\nu}(x),
\nonumber\\
&& \eta^{ac}\eta_{cb}=\delta^a_{\ b}=e^a_{\ \mu}(x)e^\mu_{\ b}(x).
\end{eqnarray}

Now, let us focus on the line element of Bonnor-Melvin-Lambda spacetime with rainbow gravity effects and later on the form of the metric (and its inverse), tetrads (and its inverses), curved gamma matrices, and spinorial and spin connections. So, starting with the Bonnor-Melvin-Lambda spacetime, we have the following line element for such a curved background \cite{Zofka,Ahmed1,Ahmed2}
\begin{equation}\label{lineelement1}
ds^2=g_{\mu\nu}(x)dx^\mu dx^\nu=dt^2-\frac{1}{2\Lambda}(dr^2+\alpha^2\sin^2 rd\theta^2)-dz^2, \ \ (\mu,\nu=t,r,\theta,z),
\end{equation}
where $0\leq r<\infty$ is the polar radial coordinate, $0\leq\theta\leq 2\pi$ is the angular or azimuthal coordinate, $-\infty<(t,z)<\infty$ are the temporal and axial coordinates, $\Lambda>0$ is the cosmological constant, and $0<\alpha\leq 1$ is a conical curvature parameter (deficit angle or constant of integration), respectively. In particular, the magnetic field (strength) for this line element is given by: $H(r)=\frac{\alpha}{\sqrt{2}}\sin r$. 

On the other hand, under the influence of rainbow gravity effects (or a framework of rainbow gravity), we must modify the temporal and spatial components in the form: $dt\to\frac{dt}{F(\xi)}$ and $dx^{i'}\to\frac{dx^{i'}}{G(\xi)}$ ($i'=r,\theta,z$), where $F(\xi)$ and $G(\xi)$ are the rainbow functions and $\xi=E/E_P=\vert E\vert/E_P$ ($0\leq\xi\leq 1$), being $E$ the relativistic total energy of the test particle and $E_P$ is the Planck energy \cite{Magueijo1,Magueijo2,Magueijo3,Amelino1,Amelino2,Bezerra2019,Bakke2018}. In that way, we can rewrite the line element \eqref{lineelement1} in terms of the rainbow functions, we obtain the following line element of the Bonnor-Melvin-Lambda spacetime with rainbow gravity effects (``Rainbow-Bonnor-Melvin-Lambda spacetime'') \cite{Ahmed3}
\begin{equation}\label{lineelement2}
ds^2(\xi)=g_{\mu\nu}(x,\xi)dx^\mu dx^\nu=\frac{dt^2}{F^2(\xi)}-\frac{1}{G^2(\xi)}\left[\frac{1}{2\Lambda}(dr^2+\alpha^2 \sin^2 r d\theta^2)+dz^2\right],
\end{equation}
where $g_{\mu\nu}(x,\xi)$ and $g^{\mu\nu}(x,\xi)$ are given by
\begin{equation}\label{metric1}
g_{\mu\nu}(x,\xi)=\left(\begin{array}{cccc}
\frac{1}{F^2(\xi)} & \ 0 & 0 & 0 \\
0 & -\frac{1}{2\Lambda G^2(\xi)} &  0 & 0 \\
0 & \ 0 & -\frac{\alpha^2 \sin^2 r}{2\Lambda G^2(\xi)} & 0 \\
0 & \ 0 & 0 & -\frac{1}{G^2(\xi)}
\end{array}\right), \ \
g^{\mu\nu}(x,\xi)=\left(\begin{array}{cccc}
F^2(\varepsilon) & \ 0 & 0 & 0 \\
0 & -2\Lambda G^2(\xi) &  0 & 0 \\
0 & \ 0 & -\frac{2\Lambda G^2(\xi)}{\alpha^2 \sin^2 r} & 0 \\
0 & \ 0 & 0 & -G^2(\xi)
\end{array}\right).
\end{equation}

However, taking the limit $E_P\gg E$ or $\xi\to 0$ (low-energy regime or infrared regime), where $F(\xi)=G(\xi)=1$, we recover the usual line element of the Bonnor-Melvin-Lambda spacetime (how should be). Besides, the standard relativistic dispersion relation is modified by rainbow gravity such as $E^2 F^2(\xi)=p^2 G^2(\xi)+m_0^2$ (``rainbow dispersion relation'').

So, with the line element well defined for our system, given by \eqref{lineelement2}, now we must define a local reference frame (laboratory frame). Consequently, we can write in this local frame the gamma matrices in a curved spacetime \cite{Oliveira7,ARXIV1,Oliveira,Lawrie}. For example, using the tetrads formalism of GR, it is possible to achieve this objective perfectly \cite{Oliveira7,ARXIV1,Oliveira,Lawrie}. In particular, this formalism states that a curved spacetime can be introduced point to point with a flat spacetime via objects of the type $e^\mu_{\ a}(x,\xi)$, which are so-called tetrads ($4\times 4$ square matrices), and which together with their inverses, given by $e^a_{\ \mu}(x,\xi)$, obey the following relations: $dx^\mu=e_{\ a}^\mu(x,\xi)\hat{\theta}^a(\xi)$ and $\hat{\theta}^b(\xi)=e^b_{\ \nu}(x,\xi)dx^\nu$, where $\hat{\theta}^a(\xi)$ ($a,b=0,1,2,3$) is a quantity so-called noncoordinate basis \cite{Oliveira7,ARXIV1,Oliveira,Lawrie}.

In this way, using the tetrads formalism, we can rewrite the line element \eqref{lineelement2} in terms of the noncoordinate basis, such as
\begin{equation}\label{lineelement3}
ds^2(\xi)=\eta_{ab}\hat{\theta}^a(\xi)\otimes\hat{\theta}^b(\xi)=(\hat{\theta}^0(\xi))^2-(\hat{\theta}^1(\xi))^2-(\hat{\theta}^2(\xi))^2-(\hat{\theta}^3(\xi))^2,
\end{equation}
where the components of $\hat{\theta}^a(\xi)$ are given by
\begin{equation}\label{bases}
\hat{\theta}^0(\xi)=\frac{1}{F(\xi)}dt, \ \ \hat{\theta}^1(\xi)=\frac{1}{\sqrt{2\Lambda}G(\xi)}dr, \ \ \hat{\theta}^2(\xi)=\frac{\alpha \sin r}{\sqrt{2\Lambda}G(\xi)}d\theta, \ \ \hat{\theta}^3(\xi)=\frac{1}{G(\xi)}dz.
\end{equation}

Therefore, this results in the following tetrads and their inverses
\begin{equation}\label{tetrads}
e^{\mu}_{\ a}(x,\xi)=\left(
\begin{array}{cccc}
 F(\xi) & 0 & 0 & 0\\
 0 & \sqrt{2\Lambda}G(\xi) & 0 & 0\\
 0 & 0 & \frac{\sqrt{2\Lambda}G(\xi)}{\alpha \sin r} & 0\\
 0 & 0 & 0 & G(\xi)\\
\end{array}
\right), \ \
e^{a}_{\ \mu}(x,\xi)=\left(
\begin{array}{cccc}
 \frac{1}{F(\xi)} & 0 & 0 & 0\\
 0 & \frac{1}{\sqrt{2\Lambda}G(\xi)} & 0 & 0\\
 0 & 0 & \frac{\alpha \sin r}{\sqrt{2\Lambda} G(\xi)} & 0\\
 0 & 0 & 0 & \frac{1}{G(\xi)}\\
\end{array}
\right).
\end{equation}

Consequently, the curved gamma matrices are given by
\begin{eqnarray}\label{gammamatrices}
&& \gamma^t(x,\xi)=F(\xi)\gamma^0,
\nonumber\\
&& \gamma^r(x,\xi)=\sqrt{2\Lambda}G(\xi)\gamma^1,
\nonumber\\
&& \gamma^\theta(x,\xi)=\frac{\sqrt{2\Lambda}G(\xi)}{\alpha \sin r}\gamma^2,
\nonumber\\
&& \gamma^z(x,\xi)=G(\xi)\gamma^3.
\end{eqnarray}

In addition, the non-null components of the Christoffel symbols are given by
\begin{eqnarray}\label{symbols}
&& \Gamma^{r}_{\ \theta\theta}(x,\xi)=-\alpha^2\sin r\cos r,
\nonumber\\
&& \Gamma^{\theta}_{\ r\theta}(x,\xi)=\Gamma^{\theta}_{\ \theta r}(x,\xi)=\frac{\cos r}{\sin r}.
\end{eqnarray}

Consequently, the non-null components of the spin connection are written as
\begin{equation}\label{spinconnection2}
\omega_{12\theta}(x,\xi)=-\omega_{21\theta}(x,\xi)=-\alpha\cos r,
\end{equation}
where implies in the following non-null component for the spinorial connection
\begin{equation}\label{spinorialconnection}
\Gamma_\theta(x,\xi)=\frac{\alpha\cos r}{2}\gamma^1\gamma^2.
\end{equation}

So, using the expressions \eqref{gammamatrices} and \eqref{spinorialconnection}, we obtain the following contribution of the spinorial connection (or spin)
\begin{equation}\label{contributionofthespinorialconnection}
\gamma^{\theta}(x,\xi)\Gamma_{\theta}(x,\xi)=\frac{\sqrt{2\Lambda}G(\xi)}{2\tan r}\gamma^1, \ \ \left(\tan r=\frac{\sin r}{\cos r}\right).
\end{equation}

In this way, we have from Eq. \eqref{dirac2} the following DE under rainbow gravity effects in the Bonnor-Melvin-Lambda spacetime
\begin{equation}\label{dirac3}
\left[iF(\xi)\gamma^0\partial_t+i\sqrt{2\Lambda}G(\xi)\gamma^1\left(\partial_r+\frac{1}{2\tan r}\right)+\frac{i\sqrt{2\Lambda}G(\xi)}{\alpha \sin r}\gamma^2\partial_\theta+iG(\xi)\gamma^3 \partial_z-m_0\right]\psi=0.
\end{equation}

Besides, here we consider a stationary quantum system where the spinor $\psi$ can be defined as follows \cite{Oliveira7,ARXIV1,Oliveira}
\begin{equation}\label{spinor}
\psi(t,r,\theta,z)=e^{i(m_j\theta+p_z z-Et)}\Phi(r), \ \ \Phi(r)=\left(
           \begin{array}{c}
            f(r) \\
            g(r) \\
           \end{array}
         \right),
\end{equation}
where $f(r)=(f_+,f_-)^T$ and $g(r)=(g_+,g_-)^T$ are two spinors with two-component each (or radial wave functions), $E$ is the relativistic total energy, $p_z=const.\geq 0$ ($-\infty<p_z<\infty$) is the module/strength of the linear momentum vector along the $z$-direction (i.e., $z$-momentum), and $m_j=\pm\frac{1}{2},\pm\frac{3}{2},\pm\frac{5}{2},\ldots$ is the total magnetic quantum number.

So, using the spinor \eqref{spinor} in \eqref{dirac3}, we obtain the following time-independent DE (or stationary DE)
\begin{equation}\label{dirac4}
\left[F(\xi)\gamma^0 E+i\sqrt{2\Lambda}G(\xi)\gamma^1\left(\frac{d}{dr}+\frac{1}{2\tan r}\right)-\frac{\sqrt{2\Lambda}G(\xi)m_j}{\alpha \sin r}\gamma^2-G(\xi)\gamma^3 p_z-m_0\right]\Phi(r)=0.
\end{equation}

Now, using the gamma matrices \eqref{Diracmatrices}, we can obtain from \eqref{dirac4} two coupled first-order differential equations for $f(r)$ and $g(r)$, given as follows
\begin{equation}\label{dirac5}
(m_0-F(\xi)E)f(r)=\left[i\sqrt{2\Lambda}G(\xi)\sigma^1\left(\frac{d}{dr}+\frac{1}{2\tan r}\right)-\frac{\sqrt{2\Lambda}G(\xi)m_j}{\alpha\sin r}\sigma^2-G(\xi)\sigma^3 p_z\right]g(r),
\end{equation}
and
\begin{equation}\label{dirac6}
(m_0+F(\xi)E)g(r)=\left[-i\sqrt{2\Lambda}G(\xi)\sigma^1\left(\frac{d}{dr}+\frac{1}{2\tan r}\right)+\frac{\sqrt{2\Lambda}G(\xi)m_j}{\alpha\sin r}\sigma^2+G(\xi)\sigma^3 p_z\right]f(r).
\end{equation}

Therefore, substituting \eqref{dirac6} in \eqref{dirac5}, we obtain the following second-order differential equation for the two components of the spinor $f(r)$
\begin{equation}\label{dirac7}
\left[\frac{d^2}{dr^2}+\frac{1}{\tan r}\frac{d}{dr}-\frac{1}{2\sin^2 r}+\frac{1}{4\tan^2 r}+\frac{sm_j}{\alpha\sin r \tan r}-\left(\frac{m_j}{\alpha\sin r}\right)^2+\frac{E^2 F^2(\xi)-m_0^2-p^2_z G^2(\xi)}{2\Lambda G^2(\xi)}\right]f_s(r)=0,
\end{equation}
where parameter $s=\pm 1$ (spinorial parameter) emerged from an eigenvalue equation given by $\sigma^3 f=\pm f=sf$ (i.e., $s$ are the eigenvalues of $\sigma^3$), where $s=+1$ is for the upper component and $s=-1$ is for the lower component. However, since the spin magnetic quantum number is labeled by $m_s=\pm 1/2=\uparrow \downarrow$ (spin up and down), it implies that we can write this quantum number in terms of $s$ such as $m_s=s/2$ (i.e, $s$ can also be called ``spin parameter'').

\section{The relativistic energy spectrum\label{sec3}}

To obtain the relativistic energy spectrum (high energy spectrum), we must try to solve Eq. \eqref{dirac7}. However, it is difficult to proceed without an adequate simplification of this equation. According to Refs. \cite{Ahmed1,Ahmed2,Ahmed3}, a good simplification can be done through a first-order approximation for the trigonometric functions, that is, assume that: $\sin r\approx r$ and $\tan r \approx r$. In this way, Eq. \eqref{dirac7} becomes
\begin{equation}\label{dirac8}
\left[\frac{d^2}{dr^2}+\frac{1}{r}\frac{d}{dr}-\frac{A^2_s}{r^2}+B^2\right]f_s(r)=0,
\end{equation}
where we define
\begin{equation}\label{dirac9}
A_s\equiv\frac{1}{\alpha}\Big\vert m_j-\frac{s\alpha}{2}\Big\vert, \ \ B\equiv\sqrt{\frac{E^2 F^2(\xi)-m_0^2-p^2_z G^2(\xi)}{2\Lambda G^2(\xi)}}.
\end{equation}

In particular, Eq. \eqref{dirac8} is a second-order Bessel equation whose solution is given in the form \cite{Ahmed1,Ahmed2,Ahmed3,Bakke2018,Faizuddin,Arfken}
\begin{equation}\label{dirac10}
f_s (r)=c_1 J_{A_s}(B r)+c_2 Y_{A_s}(B r),
\end{equation}
where $J_{A_s}(B r)$ and $Y_{A_s}(B r)$ are the Bessel functions of first and second kinds, and $c_1$ and $c_2$ are arbitrary constants. So, as at the origin ($r=0$) the Bessel function of first kind is finite/regular while the Bessel function of second kind is infinite/irregular, implies that: $c_1\neq 0$ and $c_2=0$ \cite{Ahmed1,Ahmed2,Ahmed3,Bakke2018,Faizuddin,Arfken}. Therefore, the solution \eqref{dirac10} becomes
\begin{equation}\label{dirac11}
f_s (r)=c_1 J_{A_s}(B r),
\end{equation}
where $f_s(r\to 0)=0$, i.e., our first boundary condition (that is a null radial wave function at the origin). Besides, the asymptotic form (at a large distance) of the Bessel function of first kind is given by
\begin{equation}\label{dirac12}
J_{A_s}(B r)\propto\cos\left(Br-\frac{A_s\pi}{2}-\frac{\pi}{4} \right).
\end{equation}

Now, we would like to restrict the motion of fermion to a region where a hard-wall confining potential is present, i.e., some finite distance where the radial wave function is also null (this would be our second boundary condition). So, according to the Dirichlet boundary condition, the radial wave function vanishes at some finite distance (fixed radius) where $r=r_0$, that is \cite{Ahmed1,Ahmed2,Ahmed3,Bakke2018,Faizuddin}
\begin{equation}\label{dirac13}
f_s (r=r_0)=0.
\end{equation}

Consequently, this boundary condition implies that the argument of \eqref{dirac12} must be given by \cite{Ahmed1,Ahmed2,Ahmed3,Bakke2018,Faizuddin}
\begin{equation}\label{dirac14}
\left(Br_0-\frac{A_s\pi}{2}-\frac{\pi}{4}\right)=(2n+1)\frac{\pi}{2}, \ \ (n=0,1,2,\ldots),
\end{equation}
where $n$ can be seen here as a quantum number, sometimes called radial quantum number (this makes sense since this integer number arises from a radial differential equation).

Therefore, using the quantization condition \eqref{dirac14} with \eqref{dirac9}, we obtain the following relativistic energy spectrum (allowed energy levels) for a Dirac fermion under rainbow gravity effects in the Bonnor-Melvin-Lambda spacetime
\begin{equation}\label{spectrum}
E^\pm_{n,m_j,m_s}=\pm\frac{1}{F(\xi)}\sqrt{m_0^2+G^2(\xi)p^2_z+\frac{2\Lambda G^2(\xi)\pi^2}{r^2_0}\left(n+\frac{\vert m_j-m_s\alpha\vert}{2\alpha}+\frac{3}{4}\right)^2},
\end{equation}
where the positive sign ($+$) represents the positive-energy states/solutions, i.e., a particle with spin up ($m_s=1/2$) or down ($m_s=-1/2$) and whose spectrum is given by $E_{particle}=E^+>0$, and the negative sign ($-$) represents the negative-energy states/solutions, i.e., an antiparticle with spin up ($m_s=1/2$) or down ($m_s=-1/2$) and whose spectrum is given by $E_{antiparticle}=-E^-=\vert E^-\vert>0$ (i.e., a particle with negative energy is actually an antiparticle with positive energy \cite{Greiner,Strange}), respectively. So, we see that the spectrum is quantized in terms of the quantum numbers $n$, $m_j$, and $m_s$ (and labeled through them), and explicitly depends on the rainbow functions $F(\xi)$ and $G(\xi)$ (with $\xi=\xi_{n,m_j,m_s}$), curvature parameter $\alpha$, cosmological constant $\Lambda$, fixed radius $r_0$, and on the rest energy $m_0$, and $z$-momentum $p_z$. In particular, the spectrum is intricately shaped by the underlying geometry, as characterized by the parameters $\alpha$ and $\Lambda$. Besides, it is also interesting to analyze the spectrum according to the values of the quantum numbers $m_j$ and $m_s$. Once this is done, we obtain Table \eqref{tab1}, which shows four possible configurations for the spectrum. According to this table, we see that for $m_j>0$ with $m_s=-1/2$ (config. 2), and for $m_j<0$ with $m_s=+1/2$ (config. 3), the spectrum is the same, that is, the spectrum of a particle/antiparticle with positive angular momentum and spin down (negative spin) is exactly the same as the spectrum of a particle/antiparticle with negative angular momentum and spin up (positive spin). Similarly, the spectrum of a particle/antiparticle with positive angular momentum and spin up (config. 1) is exactly the same as the spectrum of a particle/antiparticle with negative angular momentum and spin down (config. 4). In this way, we see that the spectrum with the highest values is that of configurations 2 and 3.
\begin{table}[h]
\centering
\begin{small}
\caption{Spectrum depend on the values of $m_j$ and $m_s$.} \label{tab1}
\begin{tabular}{ccc}
\hline
Configuration & $(m_j,m_s)$ & Spectrum \\
\hline
1& $(m_j>0,m_s=+1/2)$ & \ \ \ $E^\pm_{n,m_j,m_s}=\pm\frac{1}{F(\xi)}\sqrt{m_0^2+G^2(\xi)p^2_z+\frac{2\Lambda G^2(\xi)\pi^2}{r^2_0}\left(n+\frac{\vert m_j-\frac{\alpha}{2}\vert}{2\alpha}+\frac{3}{4}\right)^2}$\\
2& $(m_j>0,m_s=-1/2)$ & \ \ \ $E^\pm_{n,m_j,m_s}=\pm\frac{1}{F(\xi)}\sqrt{m_0^2+G^2(\xi)p^2_z+\frac{2\Lambda G^2(\xi)\pi^2}{r^2_0}\left(n+\frac{\vert m_j+\frac{\alpha}{2}\vert}{2\alpha}+\frac{3}{4}\right)^2}$\\
3& $(m_j<0,m_s=+1/2)$ & \ \ \ $E^\pm_{n,m_j,m_s}=\pm\frac{1}{F(\xi)}\sqrt{m_0^2+G^2(\xi)p^2_z+\frac{2\Lambda G^2(\xi)\pi^2}{r^2_0}\left(n+\frac{\vert \vert m_j\vert+\frac{\alpha}{2}\vert}{2\alpha}+\frac{3}{4}\right)^2}$\\
4& $(m_j<0,m_s=-1/2)$ & \ \ \ $E^\pm_{n,m_j,m_s}=\pm\frac{1}{F(\xi)}\sqrt{m_0^2+G^2(\xi)p^2_z+\frac{2\Lambda G^2(\xi)\pi^2}{r^2_0}\left(n+\frac{\vert \vert m_j\vert-\frac{\alpha}{2}\vert}{2\alpha}+\frac{3}{4}\right)^2}$\\
\hline
\end{tabular}
\end{small}
\end{table}

Now, let us graphically analyze the behavior of the spectrum for the three rainbow gravity scenarios as a function of the cosmological constant $\Lambda$, fixed radius $r_0$, and of the curvature parameter $\alpha$ for three different values of $n$ with $m_j=1/2$ (here, it is understood that $m_j$ plays the same role as $n$, that is, the spectrum increases with the increase of both $n$ and $m_j$). To make such an analysis, we consider (for simplicity) the spectrum of the configuration $2$ (or $3$), and only for the case of the particle ($E=E^+$). Therefore, using the three pairs of rainbow functions, given by \eqref{f1}, \eqref{f2}, and \eqref{f3}, on such a spectrum, we obtain Table \eqref{tab2}, which shows the spectrum for each one of the three scenarios, where we define $\epsilon_{n,m_j}\equiv\left[p^2_z+\frac{2\Lambda\pi^2}{r^2_0}\left(n+\frac{(m_j+\frac{\alpha}{2})}{2\alpha}+\frac{3}{4}\right)^2\right]$.
\begin{table}[h]
\centering
\begin{small}
\caption{Spectrum for each one of the three scenarios} \label{tab2}
\begin{tabular}{cc}
\hline
Scenario & Spectrum \\
\hline
First  & $E_{n,m_j}=-\frac{\eta}{2E_P}\epsilon_{n,m_j}+\sqrt{\left(\frac{\eta}{2E_P}\epsilon_{n,m_j}\right)^2+m_0^2+\epsilon_{n,m_j}}$\\
Second & $E_{n,m_j}=\frac{E_P}{\eta}\ln\left(1+\frac{\eta}{E_P}\sqrt{m_0^2+\epsilon_{n,m_j}}\right)$\\
Third & $E_{n,m_j}=-\frac{m_0^2\eta}{E_P\left(1-\frac{m_0^2\eta^2}{E^2_P}\right)}+\sqrt{\left(\frac{m_0^2\eta}{E_P\left(1-\frac{m_0^2\eta^2}{E^2_P}\right)}\right)^2+\frac{m_0^2+\epsilon_{n,m_j}}{\left(1-\frac{m_0^2\eta^2}{E^2_P}\right)}}$\\
\hline
\end{tabular}
\end{small}
\end{table}

\begin{figure}[!h]
\centering
\includegraphics[width=11.1cm]{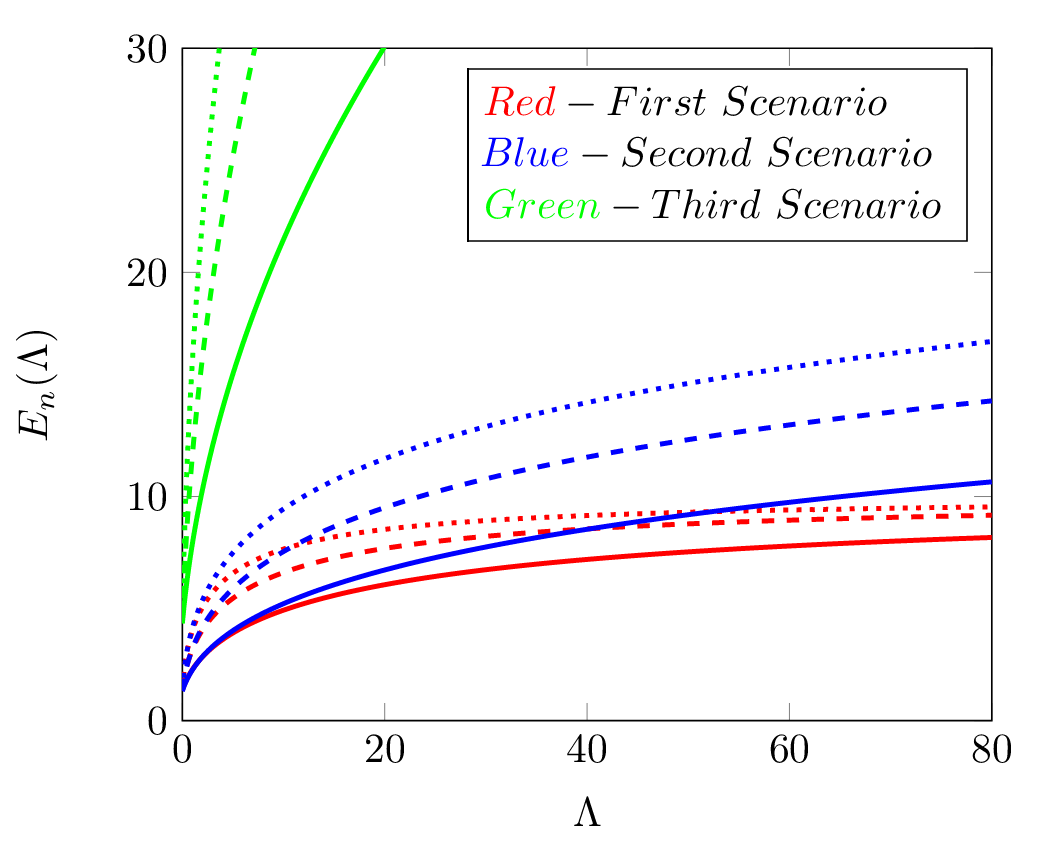}
\caption{Behavior of $E_n(\Lambda)$ vs. $\Lambda$ for three different values of $n$.}
\label{fig1}
\end{figure}

So, in Fig. \ref{fig1} we have the behavior of $E_n(\Lambda)$ vs. $\Lambda$ (energy versus cosmological constant) for three different values of $n$ (ground state, first excited state, and second excited state), where the red, blue, and green curves/lines describe the first, second, and third scenario, being $n=0$ for the solid/full curves, $n=1$ for the dashed curves, and $n=2$ for the dotted curves, respectively. To plot the graph we adopt: $m_0=p_z=1$, $\alpha=1/2$, $r_0=\pi$, $\frac{\eta}{E_P}=0.1$, and $\frac{E_P}{\eta}=10$. According to this figure, we see that the energies increase with the increase of $n$ (as it should be, otherwise, something would be wrong), that is, the energy difference between two consecutive levels is always positive ($\Delta E_n=E_{n+1}-E_n>0$). Besides, the energies increase as a function of $\Lambda$, that is, as the cosmological constant increases, the energies also increase, where the energy variation is always positive for the second and third scenario ($\Delta E(\Lambda)=E_{final}(\Lambda)-E_{initial}(\Lambda)>0$), and positive or approximately zero for the first scenario ($\Delta E(\Lambda)>0$ for $\Lambda$ small and $\Delta E(\Lambda)\approx 0$ for $\Lambda$ large). On the other hand, comparing the three scenarios, we clearly see that the energies are higher for the third scenario and lower for the first scenario (i.e., $E_0^{third}>E_0^{second}>E_0^{first}$, $E_1^{third}>E_1^{second}>E_1^{first}$, and $E_2^{third}>E_2^{second}>E_2^{first}$). However, for $\Lambda$ small (or very small) the energies of the first and second scenarios are practically equal (i.e., $E_0^{first}=E_0^{second}$, $E_1^{first}=E_1^{second}$, and $E_2^{first}=E_2^{second}$), that is, the solid, dashed, and dotted curves in red and blue coincide/overlap.

\begin{figure}[!h]
\centering
\includegraphics[width=11.1cm]{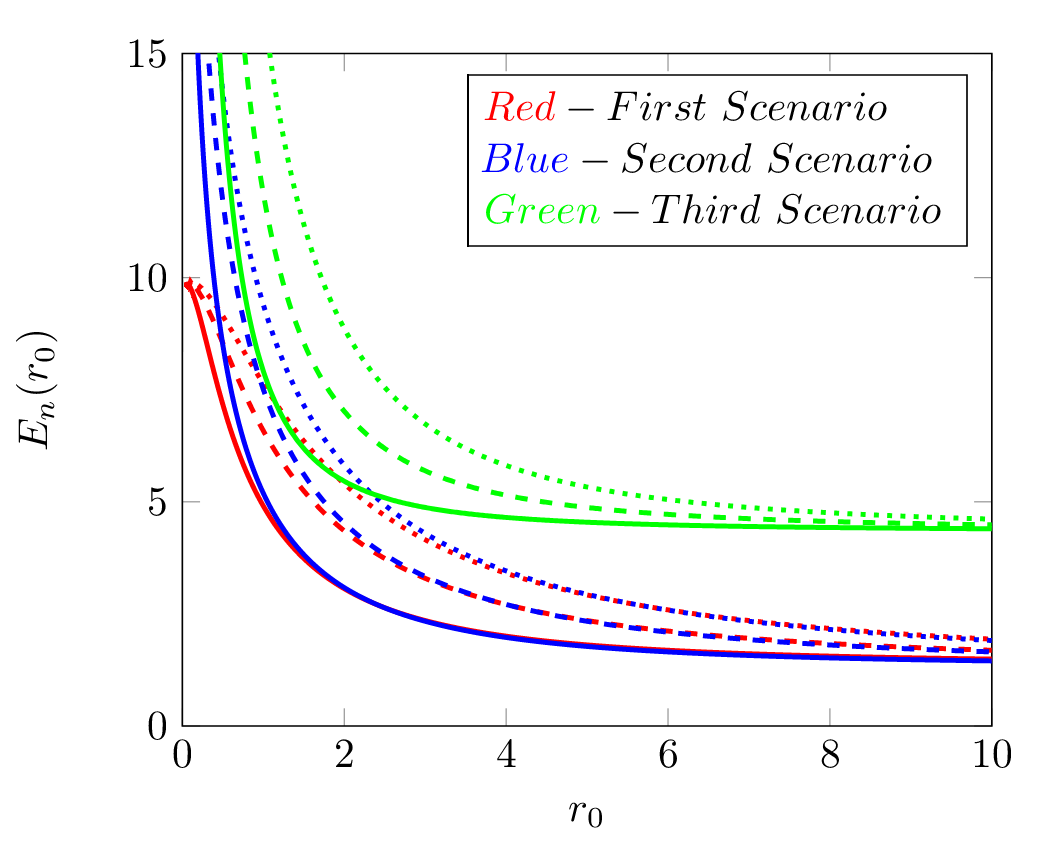}
\caption{Behavior of $E_n(r_0)$ vs. $r_0$ for three different values of $n$.}
\label{fig2}
\end{figure}

In Fig. \ref{fig2}, we have the behavior of $E_n(r_0)$ vs. $r_0$ (energy versus fixed radius) for three different values of $n$, where the red, blue, and green curves describe the first, second, and third scenario, being $n=0$ for the solid curves, $n=1$ for the dashed curves, and $n=2$ for the dotted curves, respectively. To plot the graph we adopt: $m_0=p_z=\Lambda=1$, $\alpha=1/2$, $\pi=3.14$, $\frac{\eta}{E_P}=0.1$, and $\frac{E_P}{\eta}=10$. According to this figure, we see that the energies increase with the increase of $n$ (as it should be), that is, the energy difference is positive ($\Delta E_n>0$). However, this energy difference decreases as $r_0$ increases ($\Delta E_n(r_0\to\infty)=0$), where it implies that the energy levels are sufficiently close to each other. Besides, the energies decrease as a function of $r_0$, that is, as the radius increases, the energies decrease until they remain constant at the limit $r_0\to\infty$. Therefore, the energies are greater when the particle is close to the origin and practically constant for $r_0$ large. In particular, during most of the interval of $r_0$ the energies of the first and second scenarios are practically equal (i.e., $E_0^{first}=E_0^{second}$, $E_1^{first}=E_1^{second}$, and $E_2^{first}=E_2^{second}$), that is, the solid, dashed, and dotted curves in red and blue coincide/overlap. On the other hand, comparing the three scenarios, we clearly see that the energies are also higher for the third scenario and lower for the first scenario (or second scenario since most of $r_0$ the energies of the first and second scenarios are the same).

\begin{figure}[!h]
\centering
\includegraphics[width=11.1cm]{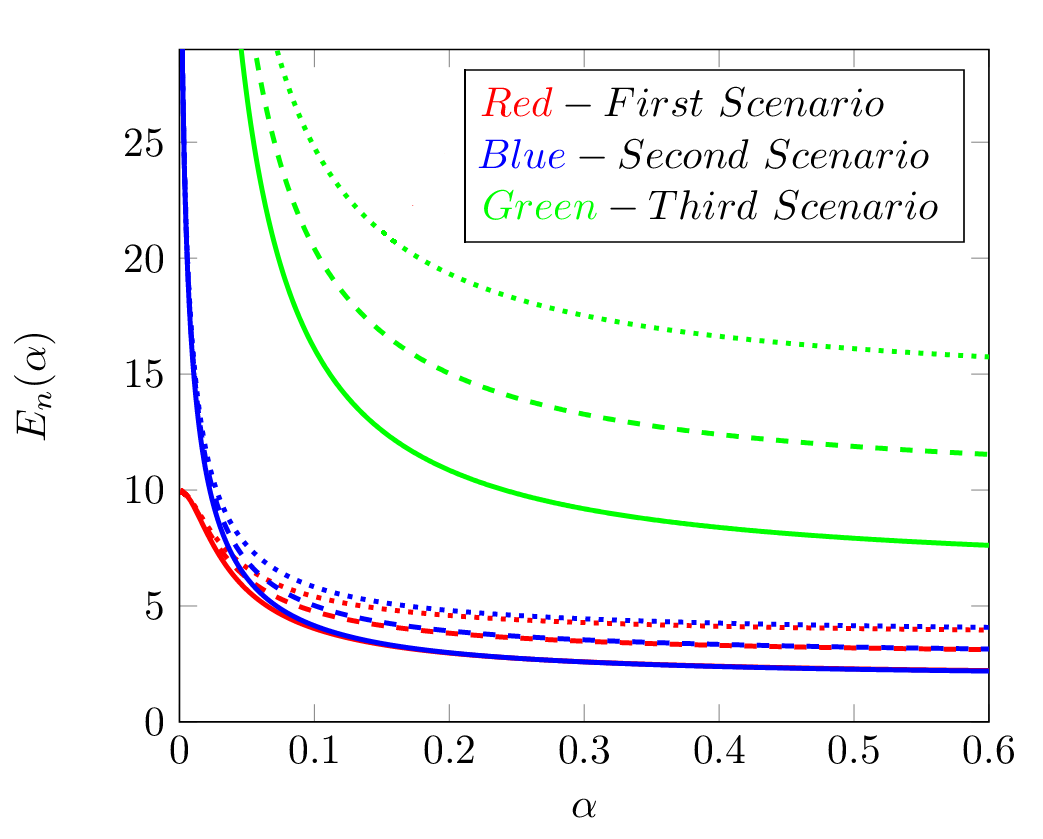}
\caption{Behavior of $E_n(\alpha)$ vs. $\alpha$ for three different values of $n$.}
\label{fig3}
\end{figure}

Already in Fig. \ref{fig3}, we have the behavior of $E_n(\alpha)$ vs. $\alpha$ (energy versus curvature parameter) for three different values of $n$, where the red, blue, and green curves describe the first, second, and third scenario, being $n=0$ for the solid curves, $n=1$ for the dashed curves, and $n=2$ for the dotted curves, respectively. To plot the graph we adopt: $m_0=p_z=\Lambda=1$, $r_0=\pi$, $\frac{\eta}{E_P}=0.1$, and $\frac{E_P}{\eta}=10$. In particular, we verified this figure is very similar to Fig. \ref{fig2}, however, it has some differences. For example, similar to Fig. \ref{fig2}, in Fig. \ref{fig3} the energies increase as $\alpha$ decreases (i.e., the energies increase with increasing curvature), during most of the interval of $\alpha$ the energies of the first and
second scenarios are practically equal, and comparing the three scenarios, we clearly see that the energies are also higher for the third scenario and lower for the first scenario. Now, unlike Fig. \ref{fig2}, in Fig. \ref{fig3} the energy difference is practically constant as $\alpha$ increases ($\Delta E_n(\alpha\to 1)=constant>0$).


\section{Conclusions}\label{sec4}

In this paper, we study the relativistic energy spectrum (high energy spectrum) for Dirac fermions under rainbow gravity effects in the $(3+1)$-dimensional Bonnor-Melvin-Lambda spacetime. To make this study, we work with the curved DE in cylindrical coordinates (or modified DE by rainbow gravity), where the formalism used was the tetrads formalism of GR. With respect to the general line element (or general metric) of our problem, we use the line element of the Bonnor-Melvin-Lambda spacetime modified by rainbow gravity, where the Bonnor-Melvin-Lambda spacetime is modeled by the cosmological constant $\Lambda$ and by a curvature parameter $\alpha$, while the rainbow gravity is modeled by two real functions: $F(\xi)$ and $G(\xi)$, being $\xi=E/E_P$ a positive real parameter (rainbow parameter), where $E$ is the relativistic total energy of the fermion and $E_P$ is the Planck energy. In particular, we consider three different pairs of rainbow functions (or three different scenarios) for $F(\xi)$ and $G(\xi)$, whose choice was based on phenomenological motivations and also because are the most worked in the literature.

So, to simplify the calculations, we consider a first-order approximation for the trigonometric functions in the second-order differential equation, where we obtain a Bessel equation. To solve this Bessel equation, we also consider a region where a hard-wall confining potential is present, i.e., some finite distance (fixed radius) where the radial wave function is null. In other words, we define a second boundary condition (Dirichlet boundary condition) to achieve the quantization of the energy. Consequently, we obtain the relativistic energy spectrum for a fermion/antifermion (or particle/antiparticle), where we discuss some characteristics of this spectrum, such as: is quantized in terms of quantum numbers $n$, $m_j$ and $m_s$, where $n=0,1,2,\ldots$ is the radial quantum number, $m_j=\pm 1/2,\pm 3/2,\pm 5/2,\ldots$ is the total magnetic quantum number, $m_s=s/2=\pm 1/2$ is the spin magnetic quantum number, and explicitly depends on the rainbow functions $F(\xi)$ and $G(\xi)$, curvature parameter $\alpha$, cosmological constant $\Lambda$, fixed radius $r_0$, and on the rest energy $m_0$, and $z$-momentum $p_z$. Besides, analyzing this spectrum according to the values of $m_j$ and $m_s$, we obtain a Table with four possible configurations for the spectrum. According to this table, we see that for $m_j>0$ with $m_s=-1/2$, and for $m_j<0$ with $m_s=+1/2$, the spectrum is the same, that is, the spectrum of a particle/antiparticle with positive angular momentum and spin down is exactly the same as the spectrum of a particle/antiparticle with negative angular momentum and spin up. Similarly, the spectrum of a particle/antiparticle with positive angular momentum and spin up is exactly the same as the spectrum of a particle/antiparticle with negative angular momentum and spin down.

Subsequently, we graphically analyze the behavior of the spectrum for the three scenarios as a function of the cosmological constant $\Lambda$, fixed radius $r_0$, and of the curvature parameter $\alpha$ for three different values of $n$ (i.e., ground state, first excited state, and second excited state). For example, in the graph $E_n(\Lambda)$ vs. $\Lambda$ (energy versus cosmological constant), we see that the energies increase with the increase of $n$ (the energy difference between two consecutive levels is always positive) and the energies increase as a function of $\Lambda$, that is, as the cosmological constant increases, the energies also increase. On the other hand, comparing the three scenarios, we clearly see that the energies are higher for the third scenario and lower for the first scenario. However, for $\Lambda$ small (or very small) the energies of the first and second scenarios are practically equal (the solid, dashed, and dotted curves in red and blue coincide/overlap). Already in the graph $E_n(r_0)$ vs. $r_0$ (energy versus fixed radius), we see that the energies increase with the increase of $n$ and the energies decrease as a function of $r_0$, that is, as the radius increases, the energies decrease until they remain constant at the limit $r_0\to\infty$. Therefore, the energies are greater when the particle is close to the origin and practically constant for $r_0$ large. In particular, during most of the interval of $r_0$ the energies of the first and second scenarios are practically equal (the solid, dashed, and dotted curves in red and blue coincide/overlap). On the other hand, comparing the three scenarios, we clearly see that the energies are also higher for the third scenario and lower for the first scenario (or second scenario since most of $r_0$ the energies of the first and second scenarios are the same).

Now, with respect to the graph $E_n(\alpha)$ vs. $\alpha$ (energy versus curvature parameter), we verified this graph is very similar to the graph $E_n(r_0)$ vs. $r_0$, however, it has some differences. For example, similar to $E_n(r_0)$ vs. $r_0$, the energies increase as $\alpha$ decreases (i.e., the energies increase with increasing curvature), during most of the interval of $\alpha$ the energies of the first and second scenarios are practically equal, and comparing the three scenarios, we clearly see that the energies are also higher for the third scenario and lower for the first scenario. However, unlike to $E_n(r_0)$ vs. $r_0$, the energy difference is practically constant as $\alpha$ increases.

\section*{Acknowledgments}

\hspace{0.5cm}The author would like to thank the Conselho Nacional de Desenvolvimento Cient\'{\i}fico e Tecnol\'{o}gico (CNPq) for financial support.

\section*{Data availability statement}

\hspace{0.5cm} This manuscript has no associated data or the data will not be deposited. [Author’s comment: There is no data because this is a theoretical paper based on calculations of the relativistic energy spectrum (high energy spectrum) for Dirac fermions under rainbow gravity effects in the $(3+1)$-dimensional Bonnor-Melvin-Lambda spacetime.]

\end{document}